\documentclass[aps,showpacs,twocolumn,prl,10pt,superscriptaddress]{revtex4-1}
 
\usepackage[utf8]{inputenc}  

\usepackage[T1]{fontenc}
\usepackage[english]{babel}  

\usepackage{amsmath,amssymb}
\usepackage{hyperref}
\usepackage{verbatim}
\usepackage{graphicx}
\graphicspath{{pics/}}

\usepackage{tabularx}
\usepackage{booktabs}
\usepackage{bm}
\usepackage{mathtools}

\usepackage[normalem]{ulem}

\usepackage{color}
\definecolor{dred}{rgb}{.8,0.2,.2}
\definecolor{ddred}{rgb}{.8,0.5,.5}
\definecolor{dblue}{rgb}{.2,0.2,.8}
\definecolor{dgreen}{rgb}{.2,0.7,.2}

\usepackage{ulem}

\newcommand{\Niso}{$^{14}$N\ }

\newcommand{\be}{\begin{equation}}
\newcommand{\ee}{\end{equation}}

\newcommand{\bal}{\begin{align}}
\newcommand{\eal}{\end{align}}
\newcommand{\bea}{\begin{eqnarray}}
\newcommand{\eea}{\end{eqnarray}}
\newcommand{\bpm}{\begin{pmatrix}}
\newcommand{\epm}{\end{pmatrix}}

\newcommand{\ket}[1]{\ensuremath{\left| #1 \right \rangle}}

\newcommand{\Hsim}{\ensuremath{H_{\text{sim}}}}

\newcommand{\trialstate}{\tau}

\begin{document}

\title{Quantum Simulation of Helium Hydride in a Solid-State Spin Register}

\author{Ya Wang}
\affiliation{3rd Institute of Physics, Research Center Scope and IQST,
University of Stuttgart, 70569 Stuttgart, Germany}

\author{Florian Dolde}
\affiliation{3rd Institute of Physics, Research Center Scope and IQST,
University of Stuttgart, 70569 Stuttgart, Germany}

\author{Jacob Biamonte}
\email{jacob.biamonte@qubit.org} 
\affiliation{ISI Foundation, Via Alassio 11/c, 10126 Torino, Italy}

\author{Ryan Babbush}
\affiliation{Department of Chemistry and Chemical Biology, Harvard University,
Cambridge, MA 02138 USA}
\affiliation{Google, 150 Main Street, Venice Beach, CA 90291, USA}

\author{Ville Bergholm}
\affiliation{ISI Foundation, Via Alassio 11/c, 10126 Torino, Italy}

\author{Sen Yang}
\affiliation{3rd Institute of Physics, Research Center Scope and IQST,
University of Stuttgart, 70569 Stuttgart, Germany}

\author{Ingmar Jakobi}
\affiliation{3rd Institute of Physics, Research Center Scope and IQST,
University of Stuttgart, 70569 Stuttgart, Germany}

\author{Philipp Neumann}
\affiliation{3rd Institute of Physics, Research Center Scope and IQST,
University of Stuttgart, 70569 Stuttgart, Germany}
 
\author{Al\'{a}n Aspuru-Guzik}
\affiliation{Department of Chemistry and Chemical Biology, Harvard University,
Cambridge, MA 02138 USA}

\author{James D.~Whitfield}
\affiliation{Vienna Center for Quantum Science and Technology, University of
Vienna, Department of Physics, Boltzmanngasse 5, Vienna, Austria 1090}

\author{J{\"o}rg Wrachtrup}
\email{j.wrachtrup@physik.uni-stuttgart.de}
\affiliation{3rd Institute of Physics, Research Center Scope and IQST,
University of Stuttgart, 70569 Stuttgart, Germany}

\date{\today}

\begin{abstract}
\emph{Ab initio} computation of molecular properties is one of the
most promising applications of quantum computing. While this problem
is widely believed to be intractable for classical computers,
efficient quantum algorithms exist which have the potential to vastly
accelerate research throughput in fields ranging from material science to
drug discovery. Using a solid-state quantum register realized in a
nitrogen-vacancy (NV) defect in diamond, we compute the bond
dissociation curve of the minimal basis helium hydride cation, HeH$^+$.
Moreover, we report an energy uncertainty (given our model basis) of the order of $10^{-14}$
Hartree, which is ten orders of magnitude below desired chemical precision.
As NV centers in diamond provide a robust and straightforward platform
for quantum information processing, our work provides several
important steps towards a fully scalable solid state implementation of
a quantum chemistry simulator.
\end{abstract}
\maketitle

Quantum simulation, as proposed by Feynman~\cite{Feynman1982} and
elaborated by Lloyd \cite{Lloyd1996} and many others
\cite{Wiesner1996,Zalka1998,Abrams1997,Berry2007,Georgescu2013},
exploits the inherent behavior of one quantum system as a resource to
simulate another quantum system. Indeed, there have
been several experimental demonstrations of quantum simulators in various
architectures including quantum optics, trapped ions, and ultracold
atoms~\cite{NatureInsights}. The importance of quantum simulators applied to 
electronic structure problems has been detailed in several recent
review articles including
\cite{Yung12,Kassal11,Brown10,Lu12,Georgesu13, THJ2014} and promises a
revolution in areas such as materials engineering, drug design and the
elucidation of biochemical processes. 

The computational cost of solving the full Schr{\"o}dinger equation 
of molecular systems using any known method on a classical computer scales exponentially with
the number of atoms involved.
However, it has been proposed that this calculation could be done
efficiently on a quantum computer,
with the cost scaling linearly in propagation time~\cite{Berry2007}.
There is now a growing body of work
proposing efficient quantum simulations of chemical Hamiltonians,
e.g.~\cite{Aspuru-Guzik2006,Lanyon2010,Whitfield2010,Kassal2010,CodyJones2012,Wecker2013,Toloui2013,Hastings2014}.
A general procedure to obtain molecular eigenenergies to a desired precision is:
$\left(i\right)$ mapping molecular wave functions into the computational basis,
$\left(ii\right)$ preparing the quantum simulator into an ansatz state
which is close to an eigenstate of the simulated Hamiltonian \Hsim{},
$\left(iii\right)$ encoding the energies into a relative phase by
simulating the time evolution operator $e^{-it \Hsim /\hbar}$
using quantum gates, and
$\left(iv\right)$ extracting the energies to desired precision
using a variant of the quantum phase estimation algorithm~\cite{Aspuru-Guzik2006,Lloyd1996,Kitaev1999}
{or, more recently, compressive sensing algorithms~\cite{Puentes2014}}.
Experimental realizations of the quantum simulation of electronic structure began
with the simulation of molecular hydrogen using quantum optics
\cite{Lanyon2010} and liquid state NMR
\cite{Du2010}.
Chemical simulation of reaction dynamics on
an eight-site lattice was then performed in NMR \cite{Lu2011}. A calculation of the energy of the helium
hydride cation in a photonics setup using a quantum variational eigensolver that avoids phase estimation has also been performed \cite{McClean2013}.


Nitrogen-vacancy (NV) centers in diamond offer a scalable and precise platform for quantum
simulation which 
does not suffer from signal losses as the system size
increases, and can avoid challenges such as 
the need for post-selected measurements. Progress to date has shown that such
systems are among the most accurate and most controllable candidates
for quantum information
processing~\cite{Gruber1997,Neumann2010,Robledo2011,Waldherr2014,Neumann2008,Togan2010,Dolde2013,Bernien2013,Dolde2014,Taminiau2014,Balasubramanian2009,Maurer2012}.
Milestone demonstrations include high-fidelity initialization and
readout~\cite{Gruber1997,Neumann2010,Robledo2011,Waldherr2014},
on-demand generation of
entanglement~\cite{Neumann2008,Togan2010,Dolde2013,Bernien2013,Waldherr2014,Dolde2014},
implementation of quantum
control~\cite{Jelezko2004,Dutt2007,Dolde2014}, ultra-long spin coherence time~\cite{Balasubramanian2009},
non-volatile memory~\cite{Maurer2012}, quantum error
correction~\cite{Waldherr2014,Taminiau2014}, as well as a host of
metrology and sensing experiments~\cite{Taylor2008,Doherty14}. Several
proposals to scale up the size of NV systems currently exist,
e.g.~\cite{Bermudez2011,Dolde2014}.
Building on this premise, 
this is the first study reporting the use of a solid state spin system
to simulate quantum chemistry.

\begin{figure}
  \includegraphics[width=1.0\columnwidth]{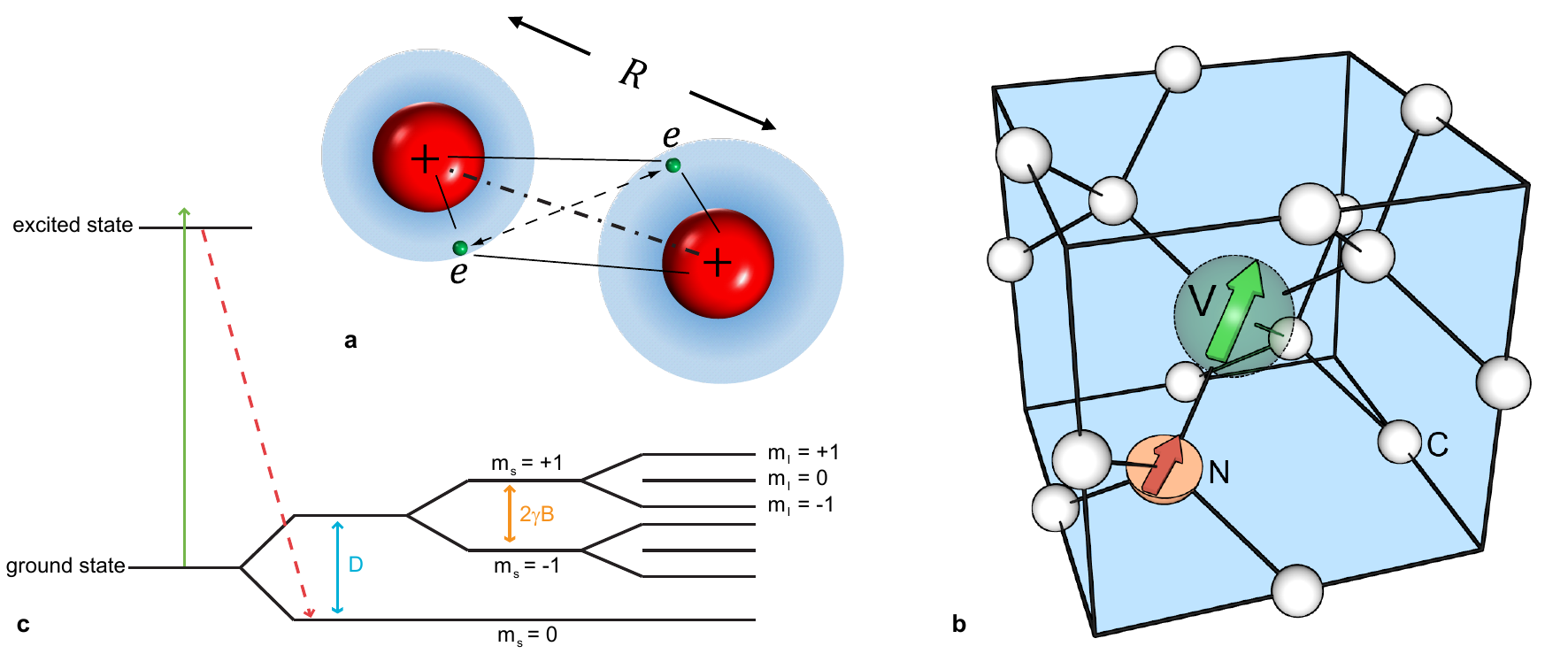}
  \caption{\label{fig:1}
    \textbf{Calculation of HeH$^+$ molecular energy with NV spin register in diamond}
		(a) HeH$^+$, molecule to be simulated. It consists of a hydrogen and a
		helium nucleus, and two electrons. The distance (bond length) between the nuclei is denoted by~$R$.
		Dot-dashed line, straight line, and dotted arrow
                indicate the nucleus-nucleus, electron-nucleus and
                electron-electron Coulomb interactions, respectively.
    (b) A nitrogen-vacancy center in diamond, used as a quantum simulator. The 
		electron spin is used for simulation and the nuclear spin
    as the probe qubit for energy readout.
    (c) Energy level diagram for the coupled spin system formed by the NV electron
			spin and nearby $^{14}$N nuclear spin. Optical transitions between ground 
			and excited state are used to
			initialize and measure the electron spin state.
    } 
\end{figure}

The chemical system we consider in this paper is the helium hydride cation,
HeH$^+$ (see Fig.~\ref{fig:1}a), believed to be the first molecule in the early universe
\cite{Engel2005}. While HeH$^+$ is isoelectronic (i.e.\ has the same
number of electrons) with the previously studied molecular hydrogen,
the reduced symmetry requires that we simulate larger subspaces of the
full configuration interaction (FCI) Hamiltonian~\Hsim{}. Specifically, we consider
\begin{equation}
	\Hsim = T_{e}+W_{ee}+V_{eN}(R)+E_{N}(R)
\end{equation}
in a minimal single particle basis with one site per atom. Here,
$T_{e}$ and $W_{ee}$ are the kinetic and Coulomb operators for the
electrons, $V_{eN}$ is the electron-nuclear interaction, and $E_{N}$
is the nuclear energy due to the Coulomb interaction between the
hydrogen and helium atoms. The last two terms depend on the
internuclear distance~$R$. 

In this work, we consider the singlet ($S=0$) sector of the electronic
Hamiltonian in a minimal single-electron basis consisting of a single
site at each atom given by contracted Gaussian orbitals.  After
taking symmetries into account, the Hamiltonian
can be represented as a $3\times 3$ matrix in the basis
$(\Psi_1$, $\Psi_6$, $\frac{1}{\sqrt{2}}\left(\Psi_3 -\Psi_4\right))$~(see Methods).
Each term of the Hamiltonian in the single particle basis
(e.g. $\langle\chi_i|(T_{e}+V_{eN})|\chi_j\rangle$) is precomputed
classically at each internuclear separation~$R$ using the canonical spin
orbitals found via the Hartree-Fock (HF) procedure which often scales as a third order polynomial in the number of basis functions.

After obtaining $\Hsim$ through this (typically) efficient classical computation, we perform the quantum simulation of
this molecule on a single-NV register, which consists of an electronic spin-1 and an associated \Niso nuclear spin-1
forming a qutrit pair (see Fig.~\ref{fig:1}b).
The electronic spin-1 of the NV system acts as the \emph{simulation register} through
mapping the molecular basis
$(\Psi_1$, $\Psi_6$, $\frac{1}{\sqrt{2}}\left(\Psi_3 - \Psi_4\right))$ onto
its $m_s = (1, 0, -1)$ states.
The \Niso nuclear spin-1 is 
used as the \emph{probe register} to read out the energies using the iterative phase estimation algorithm (IPEA)~\cite{Parker00},
as shown in Fig.~\ref{fig:1}c. 

The controlled evolution $e^{-it\Hsim }$ (we set $\hbar = 1$ from now)
on the electron spin is implemented
using
optimal control theory, which helps to realize the most precise simulation of quantum chemistry to date.
Without post-selection and at room temperature, our experimentally computed energy agrees with the corresponding classical 
calculations to within chemical precision, with a deviation of $1.4 \times10^{-14}$~Hartree.
By performing the simulation process for different values of~$R$, the electronic potential energy surfaces are
also experimentally obtained.

In order to efficiently sample the eigenenergy $E_{n}$ as the size of the system
grows, one must prepare an ansatz state that has an overlap with the corresponding eigenstate $\ket{e_{n}}$
that decreases at most polynomially in the system size.
The phase estimation algorithm~\cite{Kitaev1999} can then
be used to project the ansatz state into the exact eigenstate
with sufficiently high probability.
One possible approach to realize this requirement is to use adiabatic
state preparation~\cite{Aspuru-Guzik2006,Du2010,Biamonte2011}, the
performance of which depends on the energy gap during the entire evolution
process. An alternative approach is to approximate the eigenstate with a
trial state. Such trial states can often be prepared based on classical
approximate methods.
In our case, the simulation register
is initialized in a trial state $\ket{\trialstate} \in \{\ket{+1},\ket{-1}\}$, 
expressible as a superposition of all the \Hsim{} eigenstates, 
$\ket{\trialstate} = \sum_{k}a_{k}\ket{e_{k}}$. The probe register is
prepared in the state $\ket{\psi(0)}=(\ket{0}+\ket{-1})/\sqrt{2}$ (see Methods). 

\begin{figure}
  \includegraphics[width=1.0\columnwidth]{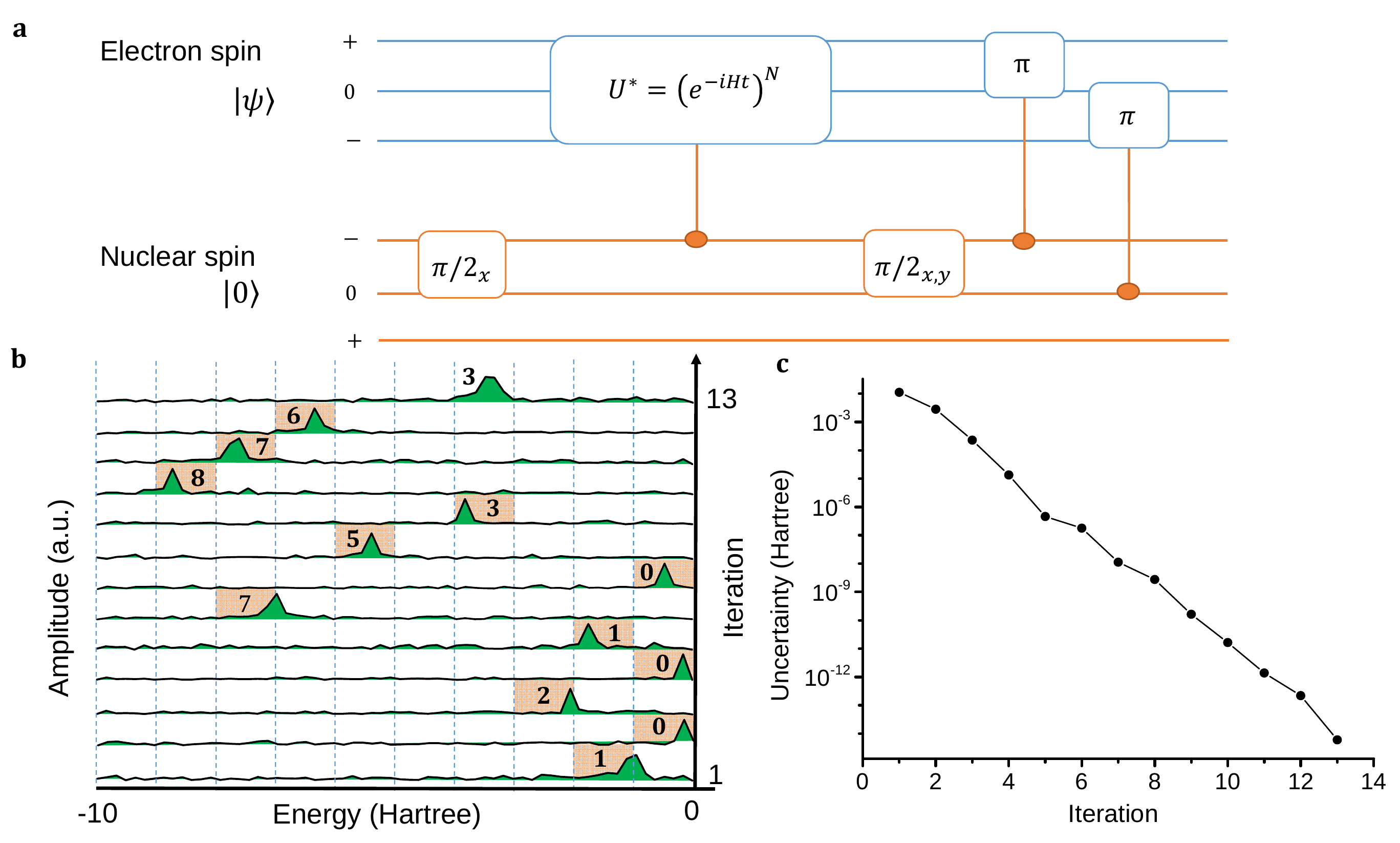}
  \caption{\label{fig:2}
    \textbf{Energy readout through quantum phase estimation algorithm}
		(a) Experimental implementation of the IPEA algorithm. The controlled gate
    $U^{*}$ is realized using optimal control (see Methods).
    The $x$, $y$ phases in the last $\pi/2$ pulse measure the real and
    imaginary parts of the signal, respectively, which yield the
    sign of the measured energy. The number of repetitions $N=10^{k-1}$ depends on
    the iteration~$k$.
    (b) Experimental results of iterative phase estimation algorithm to enhance the precision
    of measured energy for the case of $R=0.9$. The
    Fourier spectrum of the first iteration ($k=1$)
    fixes the energy roughly between $-10$ and $0$ Hartree.
    The precision is then improved iteratively by narrowing down the energy range.
    In each iteration, the energy range is
    divided into ten equal segments. The red area indicates the energy
    range for the next iteration. After each iteration at least one decimal digit, denoted by the number in
    the red area, is resolved.
    (c) The uncertainty of the measured energy
    as a function of the iteration number.}
\end{figure}

In the next step, a controlled-$U(t)$ gate for different
times~$t$, where $U(t)=\exp(-i \Hsim t)$,
is applied to encode the energies into a relative phase, resulting in the state
\begin{equation}
\ket{\psi(t)} = \frac{1}{\sqrt{2}} \sum_{k}a_{k}(\ket{0}+e^{-iE_{k}t}\ket{-1})\ket{e_{k}}.
\end{equation}
The reduced density matrix of the probe register,
\begin{equation}\label{eq:density}
\rho_{\text{probe}}(t) = \frac{1}{2}\left(
\begin{array}{cc}
 1 & \sum_{k}|a_{k}|^2e^{-iE_{k}t}  \\
 \sum_{k}|a_{k}|^2e^{iE_{k}t} & 1 \\
     \end{array}\right),
\end{equation}
contains the information about the energies in its off-diagonal elements.
This information is then transferred to the electron spin for readout by a
nuclear spin $\frac{\pi}{2}$-pulse and selective $\pi$-pulses on the electron
spin-1 (Fig.~\ref{fig:2}a). 


To measure the energy precisely, we perform classical
Fourier analysis on the signal for different times
$(t_{s}, 2 t_{s}, \ldots, L t_{s})$.
This readout method can help to resolve
the probability
$|a_{k}|^2$ of each eigenstate
$\ket{e_{k}}$ and approximate the corresponding energy~$E_k$.
We choose $t_{s}$ such that 
the sampling rate $\frac{1}{t_{s}} > |E_n|/\pi$.
To enhance the precision of the energy eigenvalues, an iterative 
phase estimation algorithm is performed. A central feature of this algorithm includes 
repeating the unitary operator $U$
to increase readout precision.
Expressing the energy as a string of decimal digits,
$E_k = x_{1}.x_{2}x_{3}\ldots$,
the first digit $x_{1}$ can be
determined by the first round phase estimation process. Once $x_{1}$ is known, the second digit $x_{2}$ 
can be iteratively determined by implementing the unitary operator $U^p$, where $p = 10$.
For the $k^\textrm{th}$ iteration, $p = 10^{k-1}$.

An increasingly precise energy can be obtained through continued iterations.
However, the repetitions and therefore the iterations are fundamentally limited
by the coherence time of the quantum system.
Moreover, the accumulated gate errors become
a dominant limitation of the energy precision as the repetitions increase. To avoid such
shortcomings, the time evolution operators $U^{p}$ are realized and optimized with optimal 
control theory (see Methods).
The precision we reach in our experiments demonstrates that 
optimal control can overcome several difficult features found when scaling up the
register size~\cite{Dolde2014}.
Although it cannot be applied in large registers to generate the
quantum gates directly, it can be used to generate flexible smaller
building blocks, ensuring high-fidelity control in future large scale
applications. 
In the present case, the method is unscalable because we compute the unitary propagator
 using a classical computer. However, by using a Trotter-type gate sequence to implement 
 the propagators, e.g.~\cite{Whitfield2010}, 
 this can be designed with polynomially scaling.


Fig.~\ref{fig:2}b shows our results of internuclear distance
$R=90$~pm with trial state $\ket{+1}$. The position of the peak
indicates the eigenvalue of molecular Hamiltonian with an offset $\operatorname{tr}(\Hsim)/3$. 
The Fourier spectrum has only one major peak, which shows that the trial state $\ket{+1}$ is close to the ground state. 
As the iterations increase, more precise decimal digits of the ground state
energy are resolved.
After 13 repetitions the molecular energy is extracted to be 
$-1.020170538763387 \pm 8\times10^{-15}$ Hartree, very close to the theoretic value, which is -1.020170538763381 Hartree, with an uncertainty of $\pm 1.4\times10^{-14}$ Hartree. 

\begin{figure}
  \includegraphics[width=0.95\columnwidth]{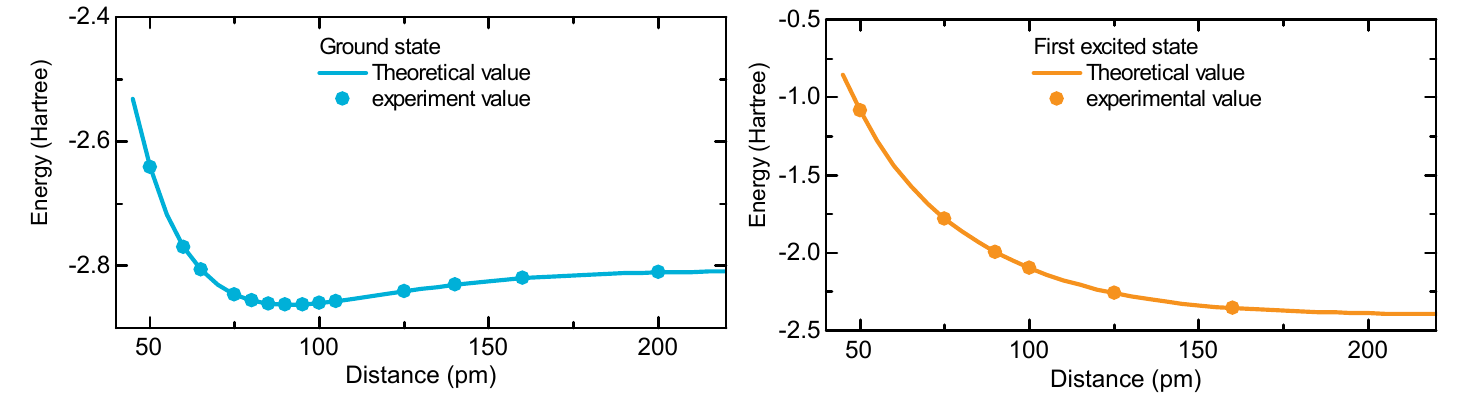}
  \caption{\label{fig:3}
    \textbf{Energy surfaces of the HeH$^+$ molecule.}
    The energy surface of the second excited state
    {can
    be obtained by subtracting energies of the the ground and first
    excited states from the trace of \Hsim{}, and}
    is not shown.
    All the measured energies are obtained in five iterations.
  }
\end{figure}

Once the energies have been measured, we can obtain the potential energy surface
of the molecule by repeating the procedure for different distances~$R$ (see Fig.~\ref{fig:3}).
The ground state energy surface is obtained with trial state $\ket{+1}$ and first excited 
state energy surface is obtained with trial state $\ket{-1}$. 
We obtain the remaining eigenenergy (of the second excited state) without further 
measurement by subtracting the ground and first excited state energies
from the trace of~\Hsim{}. 
The
potential energy surfaces can be used to compute key molecular
properties such as ionization energies and vibrational energy levels.
An important example is the equilibrium geometry: we found the minimal
energy for the ground state, $-2.86269$ Hartree, at a bond length of $91.3$~pm. 
In addition, we obtained a binding
energy of 0.07738 Hartree in our basis. 
To improve the accuracy of our results we would need to simulate the system in a larger basis, thereby requiring more qutrits.

\section{Discussion}
We will now briefly discuss several of the implications of this study.
Current quantum simulations cannot outperform classical devices.  
In large systems, the simulated propagators can be implemented using Trotter sequences and should be
accompanied by error correction. Optimal control methods, as we have demonstrated here, should prove necessary to perform these tasks with satisfactory precision.
We have demonstrated the most precise quantum simulation of
molecular energies to date,
which represents an important step towards the advanced level of
control required by future quantum simulators that will outperform
classical methods.
The energies we obtained for the helium hydride cation surpass
chemical precision by 10 orders of magnitude (with respect to the basis).
The accuracy of our results can be increased by using a larger, more
flexible single-particle basis set but this will require a larger
quantum simulator that eventually will require error correction schemes~\cite{CodyJones2012}.

 Our study presents evidence that quantum
simulators can be controlled well enough to recover increasingly
precise data.  The availability of highly accurate energy eigenvalues
of large molecules is presently far out of reach of existing
computational technology, and quantum simulation could open the door to a vast range
of new technological applications.  The approach we took was based on
iterative phase estimation~\cite{Parker00} and optimal control
decompositions~\cite{Dolde2014}---these
will form key building blocks for any solid-state quantum simulator.
Even more generally, this study would suggest that the techniques presented here should be employed in any future simulator that will outperform classical simulations of electronic structure calculations. 
 


\section{Acknowledgements}
V.B.~and J.D.B.~acknowledge financial support by Fondazione Compagnia di San Paolo
through the Q-ARACNE project. J.D.B.~would also like to acknowledge the
Foundational Questions Institute (under grant FQXi-RFP3-1322) for financial
support.
R.B.~and A.A.-G.~acknowledge support from the Air Force Office of
Scientific Research under contract FA9550-12-1-0046, as well as the National
Science Foundation CHE-1152291 and the Corning Foundation.
J.W.~acknowledges support by the EU via IP SIQS and the ERC grant SQUTEC 
as well as the DFG via the research group 1493 and SFB/TR21 and the Max Planck Society. J.D.W.~thanks the VCQ and Ford postdoctoral fellowships for support.    
We thank Mauro Faccin and Jacob Turner for providing valuable feedback
regarding the manuscript.

\bibliography{nvchem}

\newpage 

\onecolumngrid
\appendix

\section{Methods}

\subsection{Computation of molecular Hamiltonians}
The full configuration interaction Hamiltonian is a sparse matrix and each matrix element can be computed in polynomial time.
The $N$-electron Hamiltonian is asymotpically sparse. For a basis set with $M$ orbitals, there are $M^4$ terms in the Hamiltonian but the Hamiltonian is of size $\frac{M!}{N! (M-N)!}\approx M^N$ which is exponential as the number of electrons grow. 
To generate the Hamiltonian, we fix the nuclear configuration and then compute the
necessary one- and two-body integrals which parameterize the FCI
matrix at each fixed bond length
in the standard STO-3G basis~\cite{Hehre69},
using the PSI3 electronic structure package~\cite{psi3}.
 The minimal basis HeH$^+$ system has two spatial orbitals which we
denote as $g\left(r\right)$ and $e\left(r\right)$  and two spin
functions denoted as $\alpha\left(\sigma\right)$ and
$\beta\left(\sigma\right)$ which are eigenstates of the $S_z$ operator. We
combine these to form four spin orbitals,
$\chi_1= g\left(r\right) \alpha\left(r\right)$,
$\chi_2 = g\left(r\right) \beta\left(\sigma\right)$,
$\chi_3 = e\left(r\right)\alpha\left(\sigma\right)$ and
$\chi_4 = e\left(r\right) \beta\left(\sigma\right)$.
There are six possible two-electron Slater determinants,
$\Psi_1=\mathcal{A}(\chi_1\chi_2)$,
$\Psi_2=\mathcal{A}(\chi_1\chi_3)$,
$\Psi_3=\mathcal{A}(\chi_1\chi_4)$,
$\Psi_4=\mathcal{A}(\chi_2\chi_3)$,
$\Psi_5=\mathcal{A}(\chi_2\chi_4)$, and
$\Psi_6=\mathcal{A}(\chi_3\chi_4)$.
More explicitly, 
\begin{equation}
\mathcal{A}(\chi_i\chi_j)  = \frac{1}{\sqrt{2}}
\left|\begin{array}{c c}
\chi_i\left(r_1\sigma_1\right) & \chi_j\left(r_1\sigma_1\right)\\
\chi_i\left(r_2\sigma_2\right) & \chi_j\left(r_2\sigma_2\right)
\end{array}
\right|.
\end{equation}
States $\Psi_1$, $\Psi_3$, $\Psi_4$, and $\Psi_6$ have total projected spin of $M_z=0$ whereas $\Psi_2$ and $\Psi_5$ have projected values of $M_z=1$ and $M_z=-1$ respectively. Only $\Psi_1$ and $\Psi_6$ are valid eigenstates of the total spin operator $S^2$; however, the symmetric and antisymmetric combinations of $\Psi_3$ and $\Psi_4$ yield the $m_s=0$ triplet and an additional singlet, respectively. When a computation is requested on the singlet state, the PSI3 package computes the symmetry-adapted FCI matrix in the basis of $\Psi_1$, $\Psi_3$, $\Psi_4$ and $\Psi_6$. By combining $\Psi_3$ and $\Psi_4$ we obtained the three HeH$^+$ singlet states used in this experiment: $\Psi_1$, $\Psi_6$ and $\frac{1}{\sqrt{2}}\left(\Psi_3 - \Psi_4\right)$.

\subsection{Sample characteristics}
We use a nitrogen-vacancy center in high-purity diamond
grown by microwave-assisted chemical vapor deposition~(CVD).
The intrinsic nitrogen content of the grown crystal is below $1$~ppb and
the $^{12}$C content is enriched to $99.9\%$. Experiments are performed at
room temperature with an applied magnetic field of 11 gauss. The electron spin's
coherence times are $T_{2}^{*} \approx 80~\mu$s and $T_{2} \approx 600~\mu$s.

\subsection{NV system}
\newcommand{\larmor}{\omega}

In a magnetic field $B_{0}$ aligned along the NV symmetry axis, the
electronic and nuclear spin system has the Hamiltonian
\begin{align}
\label{eq:Hnv}
\notag
H/\hbar
&= 2 \pi \Delta S_z^2
+ \gamma_e B_0 S_{z} + 2 \pi A_{\text{hf}} S_{z}I_{z}
+ 2 \pi Q I_z^2 + \gamma_N B_0 I_{z}
\end{align}
where $S_{z}$ and $I_{z}$ are the dimensionless spin-1 operators for the electrons
and the \Niso{} nucleus, respectively. $\Delta \approx 2.87$~GHz and $
Q\approx -4.94$~MHz are the zero-field
splitting of the electronic spin and quadrupole splitting of the nuclear spin. The
hyperfine coupling coefficient is $A_{\text{hf}} \approx 2.16$~MHz. The Larmor
frequencies are defined as~$\larmor_i := \gamma_i B_0$, where $\gamma_i$~is the
gyromagnetic ratio of the spin (electronic or nuclear).

\subsection{System initialization}
\begin{figure}
  \includegraphics[width=1.0\columnwidth]{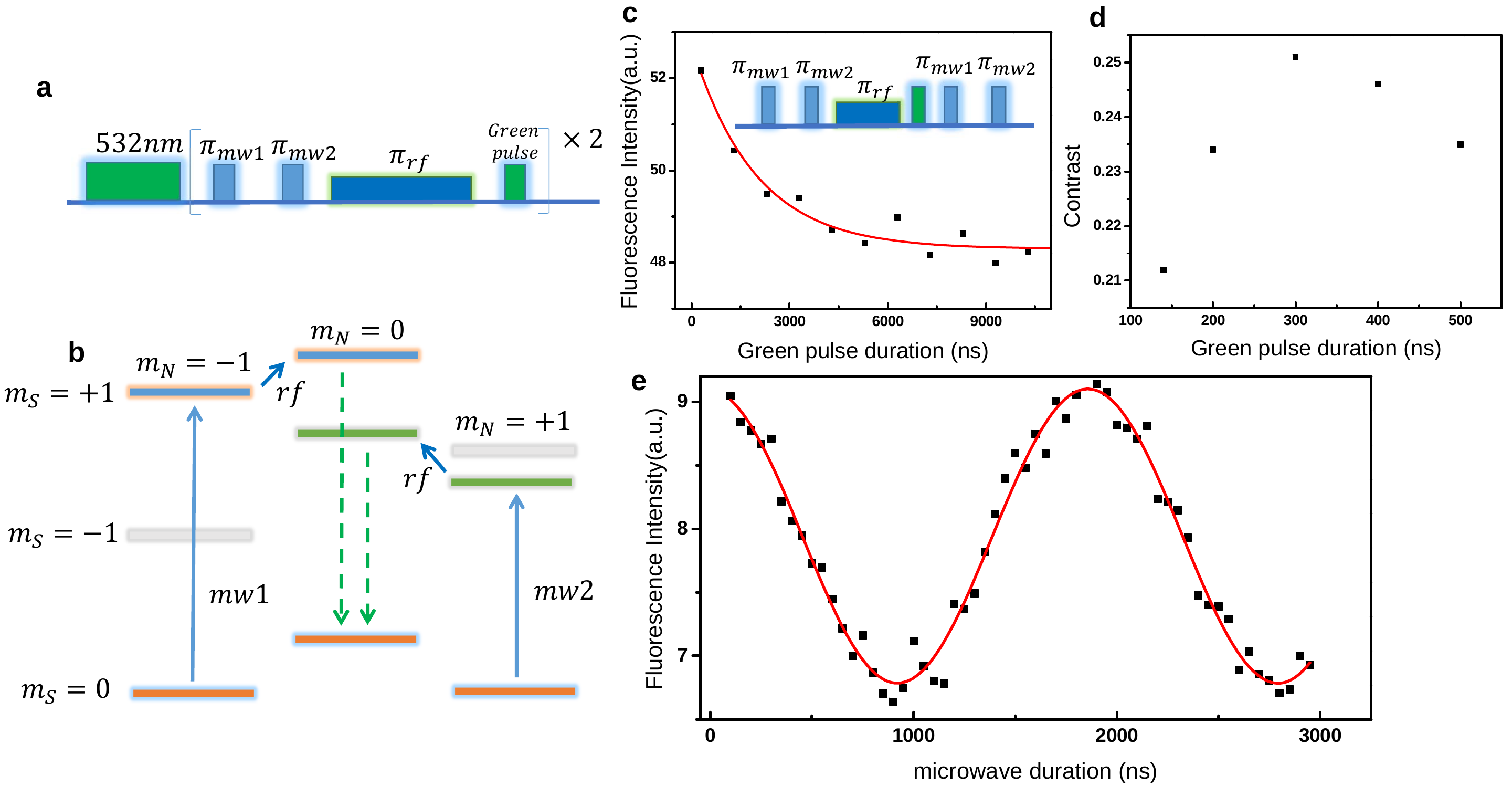}
  \caption{\label{fig:polarization}
\textbf{Polarization of \Niso nuclear spin.}
(a) 
  Experimental pulse sequence to polarize the \Niso nuclear spin. 
(b) 
 Dynamical process during one polarization step.
(c) Initialization decay of nuclear spin under laser illumination. 
The fit (red line) shows an exponential decay with time constant
$1.9\pm 0.3~\mu$s.  
The pulse sequence is shown in the inset.
(d) The contrast of electron spin Rabi oscillation in the $m_{N}=0$ subspace (here: a measure for the degree of nuclear spin initialization) varies with the second laser pulse length.
(e) The electron spin Rabi oscillation in the $m_{N}=0$ subspace
for $300$~ns green pulse duration.}
\end{figure}

In the experiment, the \Niso nuclear spin is initially in a thermal state. It
is polarized into the spin state $\ket{m_{I} = 0}$ by means of optical
pumping of the electron spin
followed polarization transfer realized with electron spin and
nuclear spin control (see Fig.~\ref{fig:polarization}). The second short laser pulse
repolarizes the electron spin into $\ket{m_{s} = 0}$, leaving the spins in the
state $\ket{m_{s} = 0, m_{I} = 0}$ \cite{Dutt2007}. In practice, the imperfect control and short
$T_{1} \approx 1.9~\mu$s time of nuclear spin under laser illumination will
result in imperfect polarization of the nuclear spin. To enhance the
polarization effect, we repeat the process two times and tune the second laser
pulse to an optimal length around $300$~ns. The observed electron spin Rabi
oscillation in the $m_{I} = 0$ subspace indicates a final polarization of
around~$60\%$. After the polarization process, the electron spin is then
prepared into the $\ket{m_{s} = +1}$ or $\ket{m_{s} = -1}$ state by another
microwave $\pi$~pulse unconditional on the nuclear spin state. Note that only the phase of the nuclear spin superposition state contains 
information in the IPEA process, therefore imperfect polarization would not affect the accuracy
of final energy measurement.

\subsection{Controlled U(t) gate realization}
In the experiment, every individual controlled gate
$U^{'}=(e^{-i \Hsim t})^{p}$
can be realized by decomposing it into more basic but highly 
complicated microwave pulses. However, this approach will
accumulate considerable control errors. 
To avoid such shortcomings, we use an alternative method,
optimal control, which has recently been used to achieve high-fidelity
control in coupled NV centers in diamond~\cite{Dolde2014}. 

To make the calculation feasible, another equivalent controlled gate
$U^{*}=e^{-iH't}$ with 
the Hamiltonian
$H' = \Hsim -\operatorname{tr}(\Hsim)/3$ is calculated. This operation will
only introduce 
additional $O(1)$ complexity. One then needs to add
this constant value $\operatorname{tr}(\Hsim)/3$ back 
to the final measured energies.

To calculate $U^{*}$, we use the GRAPE algorithm~\cite{Khaneja2005} to optimize the
pulse sequence, with the final fidelity always larger than 0.99. For every
controlled gate, the pulse sequence consists of 10 pieces of $140$~ns each.
Two microwave frequencies are applied simultaneously to control the
electron spin, in the observed hyperfine peaks of the
$\ket{m_{I} = -1,m_{s} = 0}\to\ket{m_{I} = -1 ,m_{s} = +1}$ and
$\ket{m_{I} = -1,m_{s} = 0}\to\ket{m_{I} = -1, m_{s} = -1}$ transitions.
More details about the optimal control method
can be found in reference~\cite{Dolde2014}.

\subsection{A symmetry of the ground state energy problem}

If we write the system Hamiltonian as $H = T + K$ where diagonal $T$
accounts for the HF approximations and off-diagonal $K$ accounts from
the Born-Oppenheimer approximate treatment of the problem.  We note
that whenever the support of $K$ corresponds to the adjacency matrix
of a bipartite graph, then $H = T + K$ and $L = T - K$ are
cospectral. This follows from the proof \cite{Z13} that any bipartite
(necessarily time-inversion symmetric) Hamiltonian $H$ is on the same
orbit as $-H$ under conjugation by diagonal unitarians (e.g.~there exists a diagonal unitary $\Lambda$ such that $\Lambda H \Lambda^\dagger = -H$) where $T$ is
central under this action.  Hence, they represent the same ground-state energy problems,
providing an equivalent problem instance $L$ to attempt state
preparation on. It turns out that all of the quantum chemistry
algorithms realized to date \cite{Lanyon2010, Du2010, Lu2011} have
this property including our own demonstration, where the underlying
graph corresponds to a tree.  This observation provides a second
benchmark to be considered in future experiments. 

\newpage 

\newpage 

\newpage 

\newpage 

\end{document}